# Academic Engagement and Commercialization in an Institutional Transition Environment: Evidence from Shanghai Maritime University

Dongbo Shi[1], Yeyanran Ge[1]

[1]School of International and Public Affairs,

Shanghai Jiao Tong University

**Abstract:**

Does academic engagement accelerate or crowd out the commercialization of university knowledge? Research on this topic seldom considers the impact of the institutional environment, especially when a formal institution for encouraging the commercial activities of scholars has not yet been established. This study investigates this question in the context of China, which is in the institutional transition stage. Based on a survey of scholars from Shanghai Maritime University, we demonstrate that academic engagement has a positive impact on commercialization and that this impact is greater for risk-averse scholars than for other risk-seeking scholars. Our results suggest that in an institutional transition environment, the government should consider encouraging academic engagement to stimulate the commercialization activities of conservative scholars.

## 1. Introduction

Universities are central to the public research system, and the commercial exploitation of knowledge created in the university has become increasingly vital in the world, particularly for its role in stimulating economic growth. The commercialization of academic knowledge includes activities such as disclosure, patenting, licensing, and the transfer of patents and academic entrepreneurship activities. Indeed, commercialization essentially represents the "third mission" of universities and is encouraged in many countries (Foray & Lissoni, 2010).

However, commercialization was not always as legitimate as it currently appears. The commercialization activities were initially risky and constrained by the institutional environment. In the 1960s, the attitudes of US scientists toward commercialization were ambiguous because of the trade-off between the profits of commercialization activities and the risks of political embarrassment. Most scholars had not engaged in academic commercialization until the *Bayh–Dole Act* was enacted in 1980 (Hausman, 2018).The Act shifted the attitudes of US institutions toward encouraging commercialization, and today, commercialization is regarded as one of the missions of US universities. Universities such as Stanford and the MIT are famous for their commercialization outcomes, and many universities have established technology transfer offices (TTOs) to provide services for their professors (Siegel, Waldman, Atwater, & Link, 2004).

Recent studies have highlighted the importance of academic engagement, a "knowledge-related collaboration between academic researchers with non-academic organizations" (Perkmann et al., 2013) with the aim of transferring scientific knowledge from universities to industries. Academic engagement refers to activities such as collaborative research, contract research, consulting, providing ad hoc advice and networking with practitioners. Compared to academic commercialization, academic engagement is more common and acceptable among scientists, as scholars have no need to be concerned about the legitimacy and uncertain impact of academic engagement on their scientific careers. However, the relationship between academic engagement and commercialization remains open for investigation (Perkmann et al., 2013). On the one hand, academic engagement may stimulate commercialization, because collaborating with industry can enlighten researchers about which scientific findings have more potential for commercialization, enabling them to invest more effort into patenting those findings for further licensing or spin-offs (Shane, 2001). On the other hand, academic engagement may crowd out commercialization efforts. Through academic engagement, companies absorb enough academic knowledge for their business, and researchers and universities gain considerable payback (Cohen, Nelson, & Walsh, 2002).

This study is set in the context of China, where a formal institution for the transfer of academic knowledge is far from established. Although Chinese government encourages universities to take part in commercialization activities and launched several relevant policies in 1996, 2002 and 2015, in practice, little scientific knowledge has been commercialized. Chinese scholars still fear the risks and uncertainty involved in the progress of commercialization, which range from peer pressure to gossip and even corruption-related dangers concealed within an inchoate institutional transaction process. One notable example is the case of Fu Lin[12].

---

[1] http://www.sixthtone.com/news/1001620/tsinghua-renews-contract-of-academic-charged-with-corruption

[2] Fu Lin a talent scientist at Tsinghua University who participates in academic entrepreneurship. In 2016, Fu was accused of academic corruption and arrested due

Our study finds that academic engagement has a positive impact on commercialization and this impact is larger among risk-averse scholars. Additionally, organizational support improves commercial activities. This study contributes to the existing literatures through at least two aspects. First, the relationship between academic engagement and commercialization is deeply considered in our research. Second, individual risk preference is added to our model because risk preference has a moderating effect on the relationship between academic engagement and commercialization.

The paper is organized as follows. Section 2 presents the theoretical framework and outlines the hypotheses. Section 3 explains the methodological strategy. The main results of the empirical analysis are shown in Section 4. Finally, the conclusions are presented and discussed in Section 5.

## 2. Theoretical background and hypotheses

### 2.1 Commercialization in China

Commercialization is the process of transferring academic knowledge to commercial use through discourse, patenting, and licensing (Thursby & Thursby, 2002, 2004) or by starting a for-profit company (Ding & Choi, 2011; Etzkowitz, 1983; Shane &

---

to the unintentional misuse of scientific funding from his university during the commercialization process.

Khurana, 2003; Stuart & Ding, 2006). Commercialization can make the best use of academic knowledge and stimulate the development of the economy and society, so many researchers have paid attention to the factors that affect commercialization. Studies have also found that institutional and organizational factors have an impact on commercialization (Heinecke, 2018). "Bringing institutions into evolutionary growth theory" (Nelson, 2002) and commercialization in the institutional transition stage are therefore worth of the attention of scholars.

In China, a formal institution that encourages scholar privilege has not yet been built. Several policies and legal reforms have been launched by the central government, and an informal exploration of commercialization by universities and scholars has already begun. The most three important policies were launched in 1996, 2002 and 2015. In 1996, the "Technology Transfer Law" was enacted by the National People's Congress to encourage technology transfer, and it offered details regarding technological transfer. The reward share allocated to the inventor must be no less than 20 percent of the transfer returns. This law represents the first time that researchers' privilege was officially acknowledged, but without further information regarding how to realize this privilege. In 2002, the Ministry of Science and Technology and the Ministry of Finance published “Several Provisions on Intellectual Property Management of Research Achievements of National Research Projects”, which clarified that the property rights from public funded scientific programs are belong to the universities rather than the government. In 2015, a revision to the Technology Transfer Law was published, and it required academic scholars take at least half of profits to improve their’ incentives to transfer knowledge.

**2.2 Academic engagement and commercialization**

Academic engagement is defined as “a knowledge-related collaboration by academic researchers with non-academic organizations” (Perkmann et al., 2013, Mowery et al.,2015), Academic engagement includes activities such as collaborative research and contract research as well as consulting, providing ad hoc advice and networking with practitioners (Perkmann et al., 2013).

The relationship between academic engagement and commercialization attracts considerable interest. Academic engagement that provides scholars with opportunities to establish contact with industry will stimulate commercialization (Powell & Colyvas, 2007). Academic engagement offers scholars more industry experience, social capital and social networks, which are key elements in commercialization(Stuart & Ding, 2006). Based on data from the Natural Sciences and Engineering Research Council of Canada, Landry et al. found that when a researcher has consulting experience, the likelihood of launching a spinoff increases (Landry, Amara, & Rherrad, 2006).

On the other hand, academic engagement may crowd out commercialization. Cohen (2002) stated that consulting is more vital in transferring public research to industry than patents or licenses in most industries and that the return from academic engagement is much higher than that from commercialization (Perkmann et al., 2011). Focusing on academic engagement also limits the time available for commercialization. Ding and Choi (2011) found that a company's scientific advisor is less likely to become an academic founder.

In sum, direct commercialization is not a universal way to transfer knowledge when the formal system is not established. The relationship between academic engagement and commercialization is ambiguous and the empirical results might be a net effect. However, there have been few empirical investigations into this relationship. Hence, our first step is to test the following hypotheses:

***H1a/H1b: Academic engagement has a positive/negative impact on commercialization.***

### 2.3 Risk preference and commercialization

Risk preference is one of the factors of commercialization and it moderates the relationship between academic engagement and commercialization.

The risks involved in commercialization are caused by unclear institutions, as commercialization needs time to build a sufficient history. During the early stage of institutional transition, commercialization is not widely accepted, as in the USA in the 1960s and in China currently. Scholars may encounter risks due to gossip; some people believe that scholars who spend more time on scientific research have less time to take part in commercialization. Research has shown a negative correlation between scientific productivity and technological transfer performance (Barletta, Yoguel, Pereira, & Rodriguez, 2017). Hence, scientists are afraid of being regarded as failures in terms of their scientific research. Moreover, scientists who participate in commercialization risk losing their academic freedom. Since the different goals of researchers and industry leading scholars can influence their research paths, companies may restrict their publications of outcomes for the sake of secrecy (Derrick, 2015). Commercialization is risky when scholars misunderstand ambiguous policies. An unintentional misuse of scientific funding can be easily linked to corruption, which can destroy the reputation and career development of a scientist. Hence, without sufficient confidence to manage a successful commercialization endeavor, a conservative scholar may not choose to become involved in it.

Moreover, academic engagement can aid in realizing the commercial value of knowledge with lower risk. Therefore, risk-averse scholars may tend to treat academic engagement as substitute other than supplement to commercialization.

In summary, we propose the following hypothesis:

**H2a: Risk-seeking scholars are more likely to participate in commercialization.**

**H2b: Risk preference negatively moderates the relationship between academic engagement and commercialization.**

**2.3 Organization and commercialization**

Universities have made considerable efforts to stimulate commercialization. Some universities have policies on rewarding faculty members who are involved in technological transfer that results in more licenses (Perkmann, King, & Pavelin, 2011), and some have established TTOs (Bercovitz et al., 2001) to promote commercialization. The patent and copyright policies of organizations influence the ability of scientists to commercialize (Bercovitz & Feldman, 2006; Debackere & Veugelers, 2005), and empirical results have shown that an organization's mission to spread knowledge and the existence of a regulatory environment contribute to technology transfer (Olaya Escobar et al., 2017).

Scholars who work in an organization, where commercialization is encouraged, may participate more in commercialization activities. Moreover, risks of commercialization in such organizations are lower. So scholars in such organizations may tend to treat academic engagement as supplement other than substitute to commercialization.

Hence, we hypothesize the following:

**H3a: Scholars in organizations that pay greater attention to commercialization activities are more likely to participate in commercialization.**

**H3b: Organization attention positively moderate the relationship between academic engagement and commercialization.**

Other individual characteristics, such as age, gender and previous scientific work, influence commercialization. Among all these factors, scholars' scientific work and the accumulation of such work are most influential (Louis, Blumenthal, Gluck, & Stoto, 1989; Abreu & Grinevich, 2013; Lam, 2011; Olaya Escobar, Berbegal-Mirabent, Alegre, & Duarte Velasco, 2017).

## 3. Data and Methodology

### 3.1 Data

Our data came from Shanghai Maritime University (SMU). Non-anonymous questionnaires were distributed to faculty members by the SMU TTO. The survey was conducted in July 2017, and 106 valid responses were received out of 250 questionnaires. Information on publications, patents and research funding were collected from the TTO. SMU is a public university in Shanghai with six major disciplines: engineering, management, economics, literature, science and law. There were two main considerations regarding the use of this sample. On the one hand, SMU is a medium-scale university in China. Compared to top universities such as Peking University or Tsinghua University, SMU might be more representative of universities in the country. On the other hand, SMU has the ability and resources to commercialize. Research in the maritime industry is very strong, so many of the research outcomes at SMU have the potential for further commercialization; since SMU has invested considerable effort in technology transfer, we chose to investigate the individual and organizational impacts of this practice. For example, we examined the first academic entrepreneurship case in SMU after the reform took place[3].

### 3.2 Variables

3.2.1 Dependent Variables

The key variable in this study is **commercialization**. We use a dummy variable based on the question: "Have you engaged in any of the following activities: patent licensing, patent transfer through a contract, opening a start-up either through self-investment or using patent as shares?" The variable is coded as 1 if the respondent reported at least one type of commercialization activity. We define two variables to distinguish traditional technological transfer from academic entrepreneurship. One is traditional technological transfer (Tech_Trans)[4], which captures patent licensing and transfer activities, and the other is academic entrepreneurship

---

[3]In 2016, Professor An Bowen from Shanghai Maritime University used his own patents and inventions from contract research with enterprises to establish a startup, which was the first time in China that a professor successfully established a start-up using patents as shares.

[4] Tech_Trans: If the professor had not engaged in technology transfer, the variable was coded as 0. If the professor had participated only in patent licensing or patent transfer through a contract, the variable was coded as 1. If the professor had taken part in both kinds of transfer activities, the variable was set equal to 2.

(Ac_Entre)[5].

3.2.2 Independent Variables

**Academic Engagement.** We use total funding amount of contract research to measure the extent of academic engagement. Enterprises or government enter into contracts to solve a practical problem or provide consulting services. We collect the respondent's total amount of contract funding in the past three years.

**Risk preference.**The respondents' risk preference is assessed by the extent to which they agreed with several statements, such as "I embrace risk" and "I absolutely hate risks". We use a dummy variable to measure risk preference since different respondents may have different understanding of the statements. If a respondent's mean score for the questions is higher than the mean for the population, **risk preference is coded as 1.**

**Organization**

***Org_attention.*** We used questions such as "Is your department/school focus on technological transfer, and has it even set a target?" to measure the attention of an organization. If the answer is "yes", we code the variable as 1, 0 otherwise.

---

[5] Ac_Entre: If the professor had not engaged in academic entrepreneurship, the variable was coded as 0. If the professor had taken part in establishing a start-up either by self-investment or by using patents as shares, the variable was coded as 1. If the professor had used both methods, the variable was coded as 2.

### 3.2.3 Control Variables

Individual characteristics such as age, gender, professional title, previous amount of international publications, patents, consulting work and collaboration and successful case around are controlled.

## 3.3 Econometric specification

The hypotheses regarding the relationship between academic engagement and commercialization are tested by specifying the following econometric model:

$$Commercialization = \beta_0 + \beta_1 AE + \beta_2 RP + \beta_3 ORG + \beta_4 ORG*AE + \beta_5 RP*AE + \beta_6 Controls + \varepsilon_i, \ i=1\ldots\ldots106$$

A logistic model is estimated when the dependent variable is commercialization. When examining the subset of commercialization activities, our dependent variables are ordered (0,1,2); therefore, we use ordered logistic models.

# 4. Results

## 4.1 Descriptive statistics

Table 1 summarizes the results of the descriptive analysis. According to Table 1, 15 percent of the respondents had engaged in commercialization activities. Regarding the independent variables in the table, the average funding of contract research (2014-2017) is 412,72 yuan per professor. The mean value for risk preference was 0.642. The mean value for successful cases and organization attention are 0.642 and 0.509, respectively, which indicates that 68 professors are affected by their peers and that 54 professors claim their departments pay attention to technological transfer.

Our sample include 85 males and 21 females, and the age range is from 28 to 58 years, with an average age of 37 years. Title is categorized as middle title, vice-senior title, and senior title. The respondents publish an average of 2.9 international papers and have 1.132 patents in the last three years.

------------------------------

Insert Table 1 about here

------------------------------

Table 2 presents the correlation matrix for the variables. The correlations between the variables provide support for our hypotheses and justify further examination. Significant correlations are observed between title and age, which is consistent with the reality that one's title increases with age. Specifically, the number of contract research funding (Academic_Engagement) is significantly correlated with technology transfer/commercialization activities and linked with investing methods for commercialization. The method of using patents to transfer knowledge (Tech_tran) is correlated with the method of investing, which reflects that patents are necessary for investment. These results highlight some relationships between commercialization and academic engagement.

------------------------------

Insert Table 2 about here

------------------------------

**4.2 Commercialization and academic engagement**

The results of the econometric models are demonstrated in Table 3. Models 1, 2 and 3 present the relationship between academic engagement and commercialization. Models 4, 5 and 6 take the interaction effects into consideration to examine the moderate effect of risk preference and organization attention of academic engagement on commercialization.

Models 1, 2 and 3 show that academic engagement has a significant influence on commercialization in China. In the institutional transition context, scholars with more contract research funding are more likely to take part in commercial activities, especially those involving academic entrepreneurship. Thus, our hypothesis 1 is supported. Moreover, those with higher risk preference are more likely to engage in commercialization, thereby supporting our hypothesis 2—commercialization activities are accompanied by risks, so professors will seriously consider the impacts of commercialization before making a decision about commercialization. Moreover, institutional factors such as the organizational environment have impacts on commercialization; organizations with successful cases stimulate the transfer of patents.

Models 4, 5 and 6 reveal an interactive effect of individual risk preference, organization attention and academic engagement on commercialization. In Model 4, individual risk preference and funding for contract research are both positively related to commercialization. The interactive term of contract research funding and risk preference is negatively related to commercialization, indicating that professors who are risk-seeking will be less likely to take academic engagement into consideration to achieve commercialization. The interactive term of academic engagement and organization attention is positive related to commercialization, demonstrating that organizations who encourage commercialization are likely to stimulate personal academic engagement to realize commercialization. The variable for academic entrepreneurship is not significant in our analysis, as academic entrepreneurship represents a completely new method for achieving commercialization that has seldom been attempted and therefore needs further attention. Regarding the control variables, we find a positive correlation between sex and patents and reveal that male scientists are more likely to engage in patent licensing and transfer than female scientists. A successful case of commercialization in the surrounding environment stimulates patent transfer. Thus, our hypothesis 3 is partly supported.

In sum, a clear relationship can be seen between academic engagement and commercialization. The impacts vary according to different academic engagement methods and commercialization routes, such as patenting and licensing and academic engagement. Moreover, in the context of institutional transition in China, professors care about the risks of commercialization. Additionally, institutional transition is still occurring in relation to commercialization in China, and relevant policies are therefore not mature enough to avoid risks. Thus, professors who have a low risk preference will take part in academic engagement instead of commercialization to spread their knowledge to industry. On the other hand, professors with high risk preference have less contract research funding because they are "brave" enough to license or transfer their patents directly; thus, they sign fewer contracts with enterprises. What is more, organization attention positively moderates the relationship between academic engagement and commercialization.

------------------------------

Insert Table 3 about here

------------------------------

Additional analyses were implemented for the robustness checks. We used other control variables for these analyses, for example, time spent on scientific research substituted for the research accumulation variable, and personal university and enterprise research center experience was used to replace past academic engagement. This model revealed similar results to those observed previously. Funding for contract research projects and university-enterprise experience have positive impacts on commercialization. Risk preference is related to commercialization; high risk preference leads to commercialization. Individuals who prefer risk may not have more contract research funding. Academic engagement does not represent the necessary route to commercialization, and department attention stimulates investment. These analyses provide robust results that support our conclusions.

------------------------------

Insert Table 4 about here

------------------------------

## 5. Conclusion

Academic engagement stimulates commercialization. To empirically examine the relationship between engagement and commercialization in the context of the institutional transition, questionnaire data and other relevant data were collected from SMU. Individual risk preference, organizational impact and the interactive effect of academic engagement and risk preference were taken into consideration. We found that professors who are conservative but willing to commercialize their academic knowledge may attempt to participate in academic engagement to avoid some of the risks of commercialization. In addition, organizational encouragement stimulates commercialization.

Based on our results, scholars with more contract funding for research are more likely to commercialize, especially via academic entrepreneurship. The reason for this finding is probably that those with more funding have a higher likelihood of maintaining contact with enterprises and meeting the direct needs of enterprises; thus, they can engage in commercialization more easily. Moreover, with regard to risk preference, those who embrace risk have a higher disposition for commercialization. Since there are many risks involved in commercialization due to the uncertainty of relevant policies and immature practices in China, those who are brave or prefer risk will engage in commercialization. However, regarding the interactive effect of contract funding for research and risk preference, the results indicate that professors with high risk preference may participate less in academic engagement, while those who are conservative will participate more. Because of the uncertainty in their institutions, professors will not commercialize their knowledge directly but participate in academic engagement to maintain contact with enterprises. This phenomenon is unique in the institutional transition environment. Finally, we find that organizational environment can stimulate commercialization. Since SMU is the first to help professors engage in academic entrepreneurship in public, the organization has paid greater attention than other institutions to the impacts of improving commercialization.

Taken together, these results offer a new perspective of and enrich the theory on academic engagement and commercialization. Academic engagement is not independent; it can stimulate commercialization by providing more opportunities and stronger links with industry. Risk preference should be considered a supplement to the theory, as different risk preferences result in different paths to commercialization. The institutional environment is important in the relationship between academic engagement and commercialization, and departmental attention impacts commercialization.

The results of our study have some practical implications, for example, universities can encourage academic engagement in commercialization, such as contract research and consulting. Hence, conservative scholars can allow their knowledge to be used commercially. Moreover, universities need to make use of TTOs to ameliorate the risks of professors, encourage commercialization and enable those who truly want to transfer their knowledge to do so without fear. Organizations can also prioritize specific targets and utilize the effects of peers to stimulate commercialization. Last, we cannot avoid problems in the institutional transition period of commercialization. Hence, the government needs to more seriously consider the practice of institutional design and consider adjusting its policies.

**Table 1. Descriptive Statistics（2015-2017）**

| | Observations | Mean | SD | Min | Max |
|---|---|---|---|---|---|
| **Dependent Variables** | | | | | |
| Commercialization | 106 | 0.151 | 0.360 | 0 | 1 |
| Tech_Trans | 106 | 0.208 | 0.529 | 0 | 2 |
| Ac_Entre | 106 | 0.179 | 0.513 | 0 | 2 |
| **Independent Variables** | | | | | |
| Academic Engagement | 106 | 41.272 | 114.124 | 0 | 619.5 |
| Risk_preference | 106 | 0.642 | 0.482 | 0 | 1 |
| Successful_case | 106 | 0.642 | 0.482 | 0 | 1 |
| Org_attention | 106 | 0.509 | 0.502 | 0 | 1 |
| **Controlled Variables** | | | | | |
| Male | 106 | 0.802 | 0.400 | 0 | 1 |
| Age | 106 | 37.075 | 6.295 | 28 | 58 |
| Title | 101 | 0.495 | 0.716 | 0 | 2 |
| Number of international papers | 106 | 2.953 | 3.440 | 0 | 20 |
| Number of patents | 106 | 1.132 | 2.575 | 0 | 20 |
| Past consulting service | 105 | 0.562 | 0.499 | 0 | 1 |
| Past collaboration | 105 | 0.543 | 0.501 | 0 | 1 |

**Table 2. Correlation Table**

| | [1] | [2] | [3] | [4] | [5] | [6] | [7] | [8] | [9] | [10] | [11] | [12] | [13] |
|---|---|---|---|---|---|---|---|---|---|---|---|---|---|
| [1]Commercialization | 1 | | | | | | | | | | | | |
| [2Tech_trans | 0.48*** | 1 | | | | | | | | | | | |
| [3]Ac_Entre | 0.37*** | 0.70*** | 1 | | | | | | | | | | |
| [4] Academic Engagement | 0.39*** | 0.06 | 0.26** | 1 | | | | | | | | | |
| [5] Risk_preference | 0.15 | -0.08 | -0.05 | 0.13 | 1 | | | | | | | | |
| [6] Successful_case | 0.26** | -0.08 | 0.27 | 0.15 | -0.11 | 1 | | | | | | | |
| [7]Org_attention | 0.10 | 0.24** | 0.27** | 0.02 | -0.14 | 0.33*** | 1 | | | | | | |
| [8]Male | 0.14 | 0.15 | 0.17** | 0.17** | -0.03 | -0.12 | -0.11 | 1 | | | | | |
| [9]Age | 0.23** | -0.08 | -0.007 | 0.12 | 0.13 | 0.20 | -0.13 | 0.09 | 1 | | | | |
| [10]Title | 0.15 | -0.11 | -0.01 | 0.18 | 0.07 | 0.20** | -0.10 | 0.05 | 0.68*** | 1 | | | |
| [11] Number of international papers | 0.08 | -0.09 | -0.08 | 0.05 | 0.01 | 0.001 | -0.07 | 0.12 | -0.10 | 0.26** | 1 | | |
| [12] Number of patents | 0.24** | 0.008 | 0.004 | 0.15 | -0.11 | 0.15 | 0.06 | 0.15 | 0.02 | 0.27** | 0.61*** | 1 | |
| [13] Past consulting service | 0.27** | 0.06 | 0.05 | 0.28*** | 0.09 | 0.23** | 0.01 | 0.18* | 0.39*** | 0.37** | -0.0004 | 0.27** | 1 |
| [14] Past collaboration | 0.28** | 0.04 | 0.06 | 0.22 | 0.06 | 0.28 | 0.09 | 0.26** | 0.21** | 0.31** | 0.14 | 0.29** | 0.62*** |

*** *p<0.01,* ** *p<0.05,* * *p<0.1*

**Table 3 Academic engagement, individual preference, organizational impact and commercialization (2015-2017)**

| | (1) | (2) | (3) | (4) | (5) | (6) |
|---|---|---|---|---|---|---|
| VARIABLES | Commerciali | Tech Tra | Ac Entre | Commercia | Tech Trans | Ac Entre |
| Academc_Engag ement | 0.00474* | -0.000600 | 0.00399** | 0.0810*** | 3.87e-05 | 0.0128* |
| | (0.00244) | (0.00211) | (0.00199) | (0.0288) | (0.0243) | (0.00716) |
| Risk_preference | 1.448* | 0.245 | -0.0659 | 9.250*** | 1.530* | 0.388 |
| | (0.836) | (0.696) | (0.750) | (2.956) | (0.865) | (0.881) |
| Org_attention | 0.953 | 0.883 | 1.602** | 0.339 | 0.285 | 1.962* |
| | (0.721) | (0.548) | (0.626) | (0.931) | (0.696) | (1.145) |
| Successful_case | 1.768 | 2.330** | 17.25*** | 1.577 | 2.844** | 16.97*** |
| | (1.081) | (1.066) | (0.646) | (1.077) | (1.239) | (0.447) |
| Male | 1.334 | 1.928 | 17.82*** | 1.160 | 2.000 | 17.47*** |
| | (1.164) | (1.234) | (0.812) | (1.337) | (1.451) | (0.124) |
| Age | 0.147* | -0.0320 | 0.00685 | 0.129 | -0.0800 | 0.0219 |
| | (0.0755) | (0.0755) | (0.110) | (0.0795) | (0.0839) | (0.113) |
| Title | -1.028 | -0.715 | 0.0427 | -1.047 | -1.130* | -0.256 |
| | (0.656) | (0.599) | (0.715) | (0.752) | (0.650) | (0.823) |
| Number_of_inter national_papers | -0.0291 | -0.107 | -0.0272 | -0.00772 | -0.0972 | -0.0272 |
| | (0.130) | (0.126) | (0.164) | (0.159) | (0.130) | (0.163) |
| Number_of_pate | 0.248 | 0.111 | -0.00435 | 0.574** | 0.144 | 0.00703 |
| | (0.155) | (0.128) | (0.204) | (0.227) | (0.135) | (0.211) |
| Past_consulting_ | -0.114 | 1.095 | -0.160 | 0.430 | 1.667 | -0.107 |
| | (1.034) | (1.134) | (2.028) | (0.822) | (1.140) | (2.036) |
| Past_collaboratio | 1.204 | -0.490 | -0.0893 | 0.462 | -1.030 | -0.189 |
| | (1.094) | (1.024) | (1.596) | (1.014) | (0.956) | (1.495) |
| CommercializationXRisk | | | | -0.0818*** | -0.0348*** | -0.00718 |
| | | | | (0.0290) | (0.0120) | (0.00609) |
| CommercializationXOrg | | | | 0.00624* | 0.0329 | -0.00279 |
| | | | | (0.00368) | (0.0239) | (0.00366) |
| Constant cut1 | | 4.702* | 37.79*** | | 4.039* | 38.19*** |
| | | (2.761) | (3.705) | | (2.453) | (3.789) |
| Constant cut2 | | 6.104** | 38.99*** | | 5.768** | 39.39*** |
| | | (2.914) | (3.936) | | (2.526) | (3.893) |
| Constant | -12.35*** | | | -18.93*** | | |
| | (3.918) | | | (5.322) | | |
| Observations | 100 | 100 | 100 | 100 | 100 | 100 |

Robust standard errors in parentheses

*** p<0.01, ** p<0.05, * p<0.1

**Table 4. Robustness analysis: Academic engagement, individual preference, organizational impact and commercialization**

| VARIABLES | (1)<br>Commerc | (2)<br>Tech Tra | (3)<br>Ac Entre | (4)<br>Commerc | (5)<br>Tech Trans | (6)<br>Ac Entre |
|---|---|---|---|---|---|---|
| Academic_Engagement | 0.00509* | 0.000938 | 0.00473* | 0.0403** | 0.0216* | 0.0108 |
| | (0.00243) | (0.00176) | (0.00244) | (0.0191) | (0.0128) | (0.00992) |
| Risk_preference | 1.925** | 0.491 | 0.311 | 4.746*** | 2.020* | 0.727 |
| | (0.832) | (0.977) | (0.814) | (1.820) | (1.090) | (1.197) |
| Org_attention | 0.905 | 1.069 | 2.118* | -0.225 | 0.660 | 2.323 |
| | (0.742) | (0.663) | (1.089) | (0.795) | (0.768) | (1.555) |
| Successful_case | 2.374* | 2.268** | 17.51*** | 2.977 | 2.726*** | 16.36*** |
| | (1.299) | (1.045) | (0.440) | (1.926) | (1.030) | (0.658) |
| Uni_enterprise_collaboration | 2.791*** | 0.980 | 1.312 | 3.002*** | 0.472 | 1.066 |
| | (0.935) | (0.911) | (1.002) | (1.067) | (0.836) | (1.065) |
| Uncertainty_preference | 0.757 | 0.195 | 0.535 | 0.0108 | -0.584 | 0.366 |
| | (1.309) | (1.035) | (1.107) | (1.513) | (0.986) | (1.139) |
| Social_preference | -1.324 | -1.078 | -1.523 | -1.258 | -1.217 | -1.625 |
| | (1.153) | (0.916) | (1.219) | (1.642) | (0.948) | (1.281) |
| Male | 1.943* | 2.122* | 18.53*** | 1.735* | 2.473** | 17.50*** |
| | (1.156) | (1.137) | (0.767) | (1.039) | (1.107) | (1.111) |
| Age | 0.112 | -0.00359 | 0.0198 | 0.101 | -0.0194 | 0.0275 |
| | (0.0715) | (0.0666) | (0.0810) | (0.0760) | (0.0643) | (0.0795) |
| Title | -0.326 | -0.600 | -0.162 | -0.249 | -0.870 | -0.311 |
| | (0.600) | (0.584) | (0.566) | (0.576) | (0.669) | (0.696) |
| Scientific_work_week | -0.328 | -0.280 | -0.421 | -0.261 | -0.378 | -0.453 |
| | (0.491) | (0.371) | (0.536) | (0.521) | (0.326) | (0.550) |
| Academic_EngagementXRisk | | | | -0.0401** | -0.0317*** | -0.00496 |
| | | | | (0.0195) | (0.01000) | (0.00808) |
| Academic_EngagementXOrg | | | | 0.00740* | 0.0105 | -0.00168 |
| | | | | (0.00325) | (0.0106) | (0.00334) |
| Constant cut1 | | 4.671 | 38.22*** | | 4.648* | 36.39*** |
| | | (3.018) | (3.581) | | (2.822) | (3.561) |
| Constant cut2 | | 6.075* | 39.55*** | | 6.307** | 37.72*** |
| | | (3.131) | (3.512) | | (2.902) | (3.613) |
| Constant | -10.67** | | | -12.48** | | |
| | (5.025) | | | (6.034) | | |
| Observations | 100 | 100 | 100 | 100 | 100 | 100 |

Robust standard errors in parentheses

*** p<0.01, ** p<0.05, * p<0.1